\documentclass[prd,preprint,superscriptaddress,showpacs,byrevtex]{revtex4}
\usepackage{bm}
\def\fsl#1{\setbox0=\hbox{$#1$}           
   \dimen0=\wd0                                 
   \setbox1=\hbox{/} \dimen1=\wd1               
   \ifdim\dimen0>\dimen1                        
      \rlap{\hbox to \dimen0{\hfil/\hfil}}      
      #1                                        
   \else                                        
      \rlap{\hbox to \dimen1{\hfil$#1$\hfil}}   
      /                                         
   \fi}                                         %
\newcommand{\be}{\begin{equation}}
\newcommand{\ee}{\end{equation}}
\newcommand{\bea}{\begin{eqnarray}}
\newcommand{\eea}{\end{eqnarray}}
\newcommand{\beq}{\begin{equation}}
\newcommand{\eeq}{\end{equation}}
\newcommand{\beqs}{\begin{eqnarray}}
\newcommand{\eeqs}{\end{eqnarray}}

\begin{document}
\title{ Light-Like Wilson Line in QCD Without Path Ordering }
\author{Gouranga C Nayak }
\affiliation{ 665 East Pine Street, Long Beach, New York 11561, USA }
\date{\today}
\begin{abstract}
Unlike the Wilson line in QED the Wilson line in QCD contains path ordering.
In this paper we get rid of the path ordering in the light-like Wilson line in
QCD by simplifying all the infinite number of non-commuting terms in the SU(3)
pure gauge. We prove that the light-like Wilson line in QCD naturally emerges
when path integral formulation of QCD is used to prove factorization of soft and collinear
divergences at all order in coupling constant in QCD processes at high energy colliders.
\end{abstract}
\maketitle
\pagestyle{plain}
\pagenumbering{arabic}
\section{Introduction}
In Feynman diagrams the infrared divergences appear whenever the energy-momentum $k^\mu$ involved with the massless particle
becomes very small. Similarly the collinear divergences occur when the momenta
${\vec k}, {\vec p}$ of two massless particles become parallel in the region $0<k<<p$. Typically the soft and collinear divergences
occur in the Feynman diagrams
due to momentum integration in the quantum loop diagrams involving  massless propagators and
due to momentum integration in the Feynman diagrams involving emission/absorption of massless particles.
In quantum electrodynamics (QED) the massless particle is photon and in quantum chromodynamics (QCD) the massless particle
is gluon.
The soft and collinear divergences are severe in QCD than that in QED because massless gluons interact with each other whereas
massless photons do not interact with each other. Since massless particle is always light-like one finds that soft and collinear
divergences can be described by light-like Wilson line.

However, the physical quantities measured are all soft and collinear divergences free. Hence
it is important to prove that all the non-canceling soft and collinear divergences in the perturbative
Feynman diagrams are factorized in the definition
of the (physical) gauge invariant non-perturbative quantities in QCD such as in the definition of
the parton distribution function and fragmentation function at high energy colliders because the soft and collinear limit
corresponds to long distance regime. This is done by supplying Wilson line in the definition
of the parton distribution function and fragmentation function \cite{collins}. The factorization refers to separation of short-distance
effects from the long-distance effects in quantum field theory.

The proof of factorization theorem in QCD is very non-trivial by using the diagrammatic method of QCD
\cite{collinssterman,nayaksterman} but it is enormously simplified by using the path integral method
of QCD \cite{nayak1f,nayak2f}. The main idea behind the path integral method of QCD to prove factorization is to
study the soft and collinear behavior of
non-perturbative correlation function such as $<0|{\bar \psi}(x)\psi(x'){\bar \psi}(x'')\psi(x''')...|0>$ in QCD
due to the presence of light-like Wilson line in QCD. Note that a light-like quark with light-like four-velocity
$l^\mu$ produces SU(3) pure gauge potential
at all the time-space points $x^\mu$ except at the spatial position ${\vec x}$ transverse to the motion of the quark
at the time of closest approach \cite{collinssterman,nayakj,nayake}. The soft and collinear divergences
in Feynman diagrams in QCD can be studied by using Eikonal approximation for the propagators and vertices
\cite{collins,tucci,collinssterman,berger,frederix,nayakqed,scet1,pathorder,nayaka2,nayaka3}.
Hence due to the Eikonal approximation
for soft and collinear divergences arising from the  soft and collinear gluons interaction with the light-like quark, the light-like quark finds the
gluon field $A^{\mu a}(x)$ as SU(3) pure gauge \cite{nayak1f,nayak2f}. The U(1) pure gauge
\bea
A^\mu(x)=\partial^\mu \omega(x)
\label{pgqed}
\eea
gives the light-like Wilson line in QED
\bea
e^{ie\int_{x_i}^{x_f} dx^\mu A_\mu(x)}
\label{lwqe}
\eea
which is used to study factorization of soft and collinear divergences in QED \cite{tucci,nayakqed}. In QCD the SU(3) pure gauge
\bea
T^aA^{\mu a}(x)=\frac{1}{ig}[\partial^\mu U(x)]U^{-1}(x),~~~~~~~~~~~U(x)= e^{igT^a\omega^a(x)}
\label{gtqcd}
\eea
gives the light-like Wilson line in QCD
\bea
{\cal P}e^{igT^a\int_{x_i}^{x_f} dx^\mu A_\mu^a(x)}
\label{lwqc}
\eea
which is used to study factorization of soft and collinear divergences in QCD \cite{nayak1f,nayak2f}.
Note that, unlike the Wilson line in QED in eq. (\ref{lwqe}) which does
not contain path ordering ${\cal P}$, the Wilson line in QCD in eq. (\ref{lwqc}) contains path ordering ${\cal P}$.

In this paper we get rid of the path ordering ${\cal P}$ in the light-like Wilson line in QCD by simplifying all the
infinite number of non-commuting terms in the SU(3) pure gauge in eq. (\ref{gtqcd}). We find that the light-like Wilson line in QCD
without path ordering is given by
\bea
&&{\cal P}{\rm exp}[ig\int_{x_i}^{x_f} dx^\mu { A}^a_\mu(x)T^a]\nonumber \\
&&={\rm exp}[ig T^a[\frac{1}{2~l \cdot { D}[A(x_f)]} l \cdot  \frac{d[gA(x_f)]}{dg}]^a]\times {\rm exp}[-ig T^b[\frac{1}{2~l \cdot { D}[A(x_i)]} l \cdot  \frac{d[gA(x_i)]}{dg}]^b]\nonumber \\
\label{wlabin}
\eea
where the right hand side of the above equation does not contain path ordering ${\cal P}$. In eq. (\ref{wlabin}) the
$D_\mu^{ab}[A]$ is the covariant derivative, $l^\mu$ is the light-like four velocity
and $A^{\mu a}(x)$ is the SU(3) pure gauge in QCD, which unlike U(1) pure gauge $A^\mu(x)$
in QED, contains infinite powers of $g$ \cite{nayakj}.

Since the light-like Wilson line in QCD does not depend on the path but depends only on the end
points \cite{nayak1f,nayak2f} we find from eq. (\ref{wlabin}) that the non-abelian phase or the gauge link in
QCD without path ordering is given by
\bea
{\cal P}e^{-ig \int_{0}^{\infty} d\lambda l\cdot { A}^a(x+l\lambda)T^a }={\rm exp}[ig T^a[\frac{1}{2~l \cdot { D}[A(x)]} l \cdot  \frac{d[gA(x)]}{dg}]^a]
\label{lkjni}
\eea
which is used to study factorization of soft and collinear divergences in QCD
where the right hand side of the above equation does not contain the path ordering ${\cal P}$.

In this paper we will provide a derivation of eq. (\ref{wlabin}).

In \cite{nayak1f} we have shown that the light-like Wilson line in QCD naturally emerges when path integral formulation
is used to prove NRQCD factorization at all order in coupling constant in heavy quarkonium production.
Similarly, in \cite{nayak2f} we have shown that the light-like Wilson line in QCD naturally emerges when path integral formulation
is used to prove factorization of soft and collinear divergences of the gluon distribution function at high energy colliders
at all order in coupling constant. In this paper we will prove that the light-like Wilson line in QCD naturally emerges when
path integral formulation is used to prove factorization of soft and collinear divergences of the quark distribution function at
high energy colliders at all order in coupling constant. Hence we find that the light-like Wilson line in QCD naturally emerges
when path integral formulation of QCD is used to
prove factorization of soft and collinear divergences at all order in coupling constant in QCD processes at high energy colliders.

The paper is organized as follows. In section II we derive the light-like Wilson line in QCD without path ordering as
given by eq. (\ref{wlabin}). In section III we study the gauge transformation of the light-like Wilson line in QCD without
path ordering. In section IV we prove that the
light-like Wilson line in QCD naturally emerges when path integral formulation of QCD is used to prove factorization of soft and collinear
divergences at all order in coupling constant in QCD processes at high energy colliders. Section V
contains conclusions.

\section{ Light-like Wilson Line in QCD Without Path Ordering }

The SU(3) pure gauge in QCD is given by eq. (\ref{gtqcd})
which contains infinite number of non-commuting terms.
Simplifying all the infinite number of non-commuting terms in eq. (\ref{gtqcd}) we find that the
SU(3) pure gauge $A^{\mu a}(x)$ is given by \cite{nayakj}
\bea
A^{\mu a}(x)=\partial^\mu \omega^b(x)\left[\frac{e^{gM(x)}-1}{gM(x)}\right]_{ab}
\label{pg4}
\eea
where
\bea
M_{ab}(x)=f^{abc}\omega^c(x).
\label{mab}
\eea
Expanding the exponential in eq. (\ref{pg4}) we find
\bea
A^{\mu a} (x)=[\partial^\mu \omega^b(x)]~ [1 + \frac{g}{2!}M(x)
+\frac{g^2}{3!}M^2(x)+\frac{g^3}{4!} M^3(x)  +...]_{ab}.
\label{pureg1}
\eea
In QED the U(1) pure gauge potential produced by a point charge $e$ is linearly proportional
to the electric charge $e$ \cite{collinssterman,nayakj,nayake}, {\it i.e.} ,
\bea
\partial^\mu \omega(x) \propto e.
\label{yj}
\eea
Since $\omega(x)$ is linearly proportional to $e$ we find that $\omega^a(x)$ is linearly proportional to $g$
\cite{nayakj,nayake}. Since $\omega^a(x)$ is linearly proportional to $g$ we write
\bea
\omega^a(x) = g \beta^a(x)
\label{aa}
\eea
where $\beta^a(x)$ is independent of $g$. Using eq. (\ref{aa}) in (\ref{pureg1}) we find
\bea
\frac{1}{g}A^{\mu a} (x)=[\partial^\mu \beta^b(x)]~ [1 + \frac{g^2}{2!}N(x)
+\frac{(g^2)^2}{3!}N^2(x)+\frac{(g^2)^3}{4!} N^3(x)  +...]_{ab}
\label{abnn}
\eea
where
\bea
N_{ab}(x) = f^{abc}\beta^c(x).
\label{nab}
\eea
Multiplying $g^2N_{ab}(x)$ in eq. (\ref{abnn}) we obtain
\bea
[gN(x)A^\mu(x)]^a=[\partial^\mu \beta^b(x)]~ [g^2N(x) + \frac{(g^2)^2}{2!}N^2(x)
+\frac{(g^2)^3}{3!}N^3(x)+\frac{(g^2)^4}{4!} N^4(x)  +...]_{ab}. \nonumber \\
\label{acbn}
\eea
Adding $\partial^\mu \beta^b(x)$ in eq. (\ref{acbn}) we find
\bea
{ D}^\mu[A(x)]\beta^a(x) = [\partial^\mu \beta^b(x)]~ [1+g^2N(x) + \frac{(g^2)^2}{2!}N^2(x)
+\frac{(g^2)^3}{3!}N^3(x)+\frac{(g^2)^4}{4!} N^4(x)  +...]_{ab} \nonumber \\
\label{ad}
\eea
where
\bea
{ D}_\mu^{ab}[A(x)] = \delta^{ab} \partial_\mu + gf^{acb} A_\mu^c(x).
\label{af}
\eea
Multiplying $g^2$ in eq. (\ref{abnn}) and then taking derivative with respect to $g^2$ we obtain
\bea
\frac{1}{2g}\frac{d[gA^{\mu a} (x)]}{dg}=[\partial^\mu \beta^b(x)]~ [1 + g^2N(x)
+\frac{(g^2)^2}{2!}N^2(x)+\frac{(g^2)^3}{3!} N^3(x)  +...]_{ab}.
\label{ae}
\eea
Since right hand sides of eqs. (\ref{ad}) and (\ref{ae}) are equal we find
\bea
{ D}^\mu[A(x)]\beta^a(x) = \frac{1}{2g}\frac{d[gA^{\mu a} (x)]}{dg}.
\label{pg6}
\eea
Converting $\beta^a(x)$ to $\omega^a(x)$ by using eq. (\ref{aa}) we find from eq. (\ref{pg6})
\bea
{ D}^\mu[A(x)]\omega^a(x) = \frac{1}{2}\frac{d[gA^{\mu a} (x)]}{dg}.
\label{pureg6}
\eea
Multiplying the same $x^\mu$ independent four vector $l^\mu$ in eq. (\ref{pureg6}) we find
\bea
l \cdot \frac{d[gA^a(x)]}{dg} =2~ l \cdot { D}[A(x)] \omega^a(x).
\label{cf}
\eea
Dividing $l \cdot D[A(x)]$ from left in eq. (\ref{cf}) we obtain
\bea
\omega^a(x) = [\frac{1}{2~l \cdot { D}[A(x)]}]_{ab}  ~ \frac{d[l \cdot gA^b(x)]}{dg}= [\frac{1}{2~l \cdot { D}[A(x)]} ~l \cdot  \frac{d[gA(x)]}{dg}]^a
\label{ag}
\eea
which gives the non-abelian phase
\bea
\Phi(x)=e^{igT^a\omega^a(x)}= {\rm exp}[igT^a [\frac{1}{2~l \cdot { D}[A(x)]} ~l \cdot  \frac{d[gA(x)]}{dg}]^a].
\label{ah}
\eea
From \cite{nayak1f,nayak2f} we find that the light-like Wilson line in QCD for soft and collinear divergences is given by
\bea
{\cal P}e^{ig \int_{x_i}^{x_f} dx^\mu A_\mu^a(x)T^a }=e^{igT^a\omega^a(x_f)}e^{-igT^b\omega^b(x_i)}=\left[{\cal P}e^{-ig \int_{0}^{\infty} d\lambda l\cdot { A}^a(x_f+l\lambda)T^a }\right]{\cal P}e^{ig \int_{0}^{\infty} d\lambda l\cdot { A}^b(x_i+l\lambda)T^b }. \nonumber \\
\label{lkjn}
\eea
Using eq. (\ref{ah}) in eq. (\ref{lkjn}) we find that the light-like Wilson line in QCD without path ordering
is given by
\bea
&&{\cal P}{\rm exp}[ig\int_{x_i}^{x_f} dx^\mu { A}^a_\mu(x)T^a]\nonumber \\
&&={\rm exp}[ig T^a[\frac{1}{2~l \cdot { D}[A(x_f)]} l \cdot  \frac{d[gA(x_f)]}{dg}]^a]\times {\rm exp}[-ig T^b[\frac{1}{2~l \cdot { D}[A(x_i)]} l \cdot  \frac{d[gA(x_i)]}{dg}]^b]\nonumber \\
\label{wlabf}
\eea
which reproduces eq. (\ref{wlabin}) where the right hand side does not contain the path ordering ${\cal P}$.

Since the light-like Wilson line in QCD does not depend on the path but depends only on the end
points \cite{nayak1f,nayak2f} we find from eqs. (\ref{lkjn}) and (\ref{ah}) that the non-abelian phase or the
gauge link in QCD without path ordering is given by
\bea
{\cal P}e^{-ig \int_{0}^{\infty} d\lambda l\cdot { A}^a(x+l\lambda)T^a }={\rm exp}[ig T^a[\frac{1}{2~l \cdot { D}[A(x)]} l \cdot  \frac{d[gA(x)]}{dg}]^a]
\label{lkjnf}
\eea
which reproduces eq. (\ref{lkjni}) which is used to study factorization of soft and collinear divergences in QCD
where the right hand side of the above equation does not contain the path ordering ${\cal P}$.

\section{ Non-Abelian Gauge Transformation of light-like Wilson Line in QCD Without Path Ordering }

In order to study the gauge transformation of the light-like Wilson line in QCD without path ordering
we proceed as follows. The non-abelian gauge transformation is given by
\bea
T^aA'^a_\mu(x) = U(x)T^aA^a_\mu(x) U^{-1}(x)+\frac{1}{ig}[\partial_\mu U(x)] U^{-1}(x)
\label{gtg}
\eea
where
\bea
U(x)=e^{igT^a\omega^a(x)}.
\label{u}
\eea
Since the matrices $T^a$ are non-commuting we find from eq. (\ref{u})
\bea
&& T^aU^{-1}(x) = T^a e^{-igT^b\omega^b(x)} = T^a[1+ (-ig) T^b\omega^b(x) +\frac{(-ig)^2}{2!} T^bT^c\omega^b(x) \omega^c(x) \nonumber \\
&& + \frac{(-ig)^3}{3!} T^bT^cT^d\omega^b(x) \omega^c(x) \omega^d(x) + \frac{(-ig)^4}{4!} T^bT^cT^d T^e\omega^b(x) \omega^c(x) \omega^d(x) \omega^e(x)+ ...].
\label{ex1}
\eea
By repeated use of the commutation relation
\bea
[T^a,~T^b]=if^{abc}T^c
\label{tab}
\eea
we find from eq. (\ref{ex1})
\bea
&&T^aU^{-1}(x)= [T^a+ (-ig) T^b\omega^b(x)T^a +\frac{(-ig)^2}{2!} T^bT^c\omega^b(x) \omega^c(x) T^a\nonumber \\
&& + \frac{(-ig)^3}{3!} T^bT^cT^d \omega^b(x) \omega^c(x) \omega^d(x) T^a + \frac{(-ig)^4}{4!} T^b \omega^b(x) T^c \omega^c(x) T^d \omega^d(x) T^e \omega^e(x) T^a+...\nonumber \\
&& + (-ig)if^{abc} \omega^b(x) T^c +(-ig)^2 T^b \omega^b(x) if^{acd} \omega^c(x) T^d +\frac{(-ig)^2}{2!} if^{abd}\omega^b(x) if^{dce} \omega^c(x) T^e\nonumber \\
&& + \frac{(-ig)^3}{2!} T^b \omega^b(x) T^c \omega^c(x) if^{ade} \omega^d(x) T^e + \frac{(-ig)^3}{2!} T^b \omega^b(x) if^{ace} \omega^c(x) if^{edg} \omega^d(x) T^g \nonumber \\
&& + \frac{(-ig)^3}{3!} if^{abe} \omega^b(x) if^{ecg} \omega^c(x) if^{gdh} \omega^d(x) T^h
 + \frac{(-ig)^4}{3!} T^b \omega^b(x) T^c \omega^c(x) T^d \omega^d(x) if^{aeg} \omega^e(x) T^g \nonumber \\
&& + \frac{(-ig)^4}{2!2!} T^b \omega^b(x) T^c \omega^c(x) if^{adg} \omega^d(x) if^{geh} \omega^e(x) T^h + \frac{(-ig)^4}{3!} T^b \omega^b(x) if^{acg} \omega^c(x) if^{gdh} \omega^d(x) if^{hei} \omega^e(x) T^i \nonumber \\
&& + \frac{(-ig)^4}{4!} if^{abg} \omega^b(x) if^{gch} \omega^c(x) if^{hdi} \omega^d(x) if^{iej} \omega^e(x) T^j+ ...]
\label{ex1p}
\eea
which gives after simplification
\bea
&& T^aU^{-1}(x) = [T^a+ (-ig) T^b\omega^b(x)T^a +\frac{(-ig)^2}{2!} T^bT^c\omega^b(x) \omega^c(x) T^a\nonumber \\
&& + \frac{(-ig)^3}{3!} T^bT^cT^d \omega^b(x) \omega^c(x) \omega^d(x) T^a + \frac{(-ig)^4}{4!} T^b \omega^b(x) T^c \omega^c(x) T^d \omega^d(x) T^e \omega^e(x) T^a+...\nonumber \\
&&+[1+(-ig)T^b\omega^b(x)+ \frac{(-ig)^2}{2!} T^b \omega^b(x) T^c \omega^c(x) +...](-ig)if^{ade} \omega^d(x) T^e  \nonumber \\
&& +[1+ (-ig)T^b\omega^b(x)+\frac{(-ig)^2}{2!} T^b \omega^b(x) T^c \omega^c(x)+...]\frac{(-ig)^2}{2!} if^{apd}\omega^p(x) if^{dhe} \omega^h(x) T^e\nonumber \\
&& +[1+ (-ig)T^q\omega^q(x)+...] \frac{(-ig)^3}{3!} if^{abe} \omega^b(x) if^{ecg} \omega^c(x) if^{gdh} \omega^d(x) T^h\nonumber \\
&&+[1+...] \frac{(-ig)^4}{4!} if^{abp} \omega^b(x) if^{pch} \omega^c(x) if^{hdq} \omega^d(x) if^{qes} \omega^e(x) T^s+ ...].
\label{ggfj}
\eea
From eq. (\ref{ggfj}) we find
\bea
&& T^aU^{-1}(x)=U^{-1}(x) [T^a+(-ig)if^{ade} \omega^d(x) T^e  +
\frac{(-ig)^2}{2!} if^{agd}\omega^g(x) if^{dhe} \omega^h(x) T^e  \nonumber \\
&&+\frac{(-ig)^3}{3!} if^{abe} \omega^b(x) if^{ecg} \omega^c(x) if^{gdh} \omega^d(x) T^h \nonumber \\
&&  +\frac{(-ig)^4}{4!} if^{abg} \omega^b(x) if^{gch} \omega^c(x) if^{hdi} \omega^d(x) if^{iej} \omega^e(x) T^j+ ...]
\label{ex2}
\eea
which gives
\bea
U(x)T^aU^{-1}(x)= [e^{-gM(x)}]_{ab} T^b
\label{utuinv}
\eea
where $M_{ab}(x)$ is given by eq. (\ref{mab}). From eq. (\ref{utuinv}) we find
\bea
U(x)T^aA_\mu^a(x)U^{-1}(x)= [e^{gM(x)}]_{ab} T^aA_\mu^b(x).
\label{ut1}
\eea

Similarly by simplifying infinite number of non-commuting terms in $[\partial_\mu U(x)]U^{-1}(x)$
we find \cite{nayakj}
\bea
\frac{1}{ig}[\partial_\mu U(x)]U^{-1}(x)=[\frac{e^{gM(x)}-1}{gM(x)}]_{ab}~[\partial_\mu \omega^b(x)]T^a
\label{pppl}
\eea
where $M_{ab}(x)$ is given by eq. (\ref{mab}).

Hence by using eqs. (\ref{ut1}) and (\ref{pppl}) in eq. (\ref{gtg}) we find
\bea
{A'}_\mu^a(x) = [e^{gM(x)}]_{ab}A_\mu^b(x) ~+ ~[\frac{e^{gM(x)}-1}{gM(x)}]_{ab}~[\partial_\mu \omega^b(x)]
\label{te}
\eea
which is the finite gauge transformation in QCD where $M_{ab}(x)$ is given by eq. (\ref{mab}).
Under infinitesimal gauge transformation we find from eq. (\ref{te})
\bea
A'^{\mu a}(x)= A^{\mu a}(x) +gf^{abc} \omega^c(x) A^{\mu b}(x) + \partial^\mu \omega^a(x)
\label{infgtf}
\eea
which is the infinitesimal gauge transformation in QCD which is familiar in the literature \cite{muta}.

When $A^{\mu a}(x)$ is the SU(3) pure gauge we find by using eq. (\ref{pg4}) in (\ref{te}) that
\bea
A'^{\mu a}(x) =  [\frac{e^{2gM(x)}-1}{gM(x)}]_{ab}~[\partial^\mu \omega^b(x)].
\label{tg}
\eea
By using eq. (\ref{aa}) in (\ref{tg}) we find
\bea
A'^{\mu a}(x)  =  [\frac{e^{2g^2N(x)}-1}{gN(x)}]_{ab}~[\partial^\mu \beta^b(x)]
\label{tga}
\eea
where $N_{ab}(x)$ is given by eq. (\ref{nab}) which is independent of $g$ because $\beta^a(x)$ is independent
of $g$, see eq. (\ref{aa}). Multiplying the matrix $gN(x)$ in eq. (\ref{tga})
we obtain
\bea
{D}^\mu[A'(x)]\beta^a(x) =  [e^{2g^2N(x)}]_{ab}~[\partial^\mu \beta^b(x)]
\label{th}
\eea
where
\bea
{ D}_\mu^{ab}[A'(x)] = \delta^{ab} \partial_\mu + gf^{acb} {A'}_\mu^c(x).
\label{afp}
\eea
By multiplying $g$ in eq. (\ref{tga}) and then taking the derivative with respect to $g$ we find
\bea
\frac{d[g A'^{\mu a}(x) ]}{dg} =  4g ~[e^{2g^2N(x)}]_{ab}~[\partial^\mu \beta^b(x)].
\label{ti}
\eea
Using eq. (\ref{th}) in (\ref{ti}) we obtain
\bea
\frac{d[g A'^{\mu a}(x) ]}{dg} =  4g~{ D}^\mu[A'(x)]\beta^a(x).
\label{tj1av}
\eea
By using eq. (\ref{aa}) in (\ref{tj1av}) we find
\bea
\frac{d[g A'^{\mu a}(x) ]}{dg} =  4~{ D}^\mu[A'(x)]\omega^a(x).
\label{tj1a}
\eea

By multiplying the same $x^\mu$ independent four vector $l^\mu$ in eq. (\ref{tj1a}) we obtain
\bea
l \cdot \frac{d[g {A'}^a(x)]}{dg} =  4~l \cdot { D}[A'(x)]\omega^a(x).
\label{tj}
\eea
By dividing $l \cdot { D}[A'(x)]$ from left in eq. (\ref{tj}) we find
\bea
[\frac{1}{2~l \cdot { D}[A'(x)]}~l \cdot \frac{d[g {A}'(x)]}{dg}]^a = 2 \omega^a(x).
\label{tk}
\eea
Under the non-abelian gauge transformation as given by eq. (\ref{gtg}) we find
from eq. (\ref{ah})
\bea
\Phi'(x)= {\rm exp}[igT^a [\frac{1}{2~l \cdot { D}[A'(x)]} ~l \cdot  \frac{d[gA'(x)]}{dg}]^a].
\label{ahxx}
\eea
Hence from eqs. (\ref{ahxx}), (\ref{tk}), (\ref{ah}) and (\ref{u}) we find
\bea
\Phi'(x)=U(x) ~\Phi(x),~~~~~~~~~~~~~~~\Phi'^\dagger(x)= \Phi^\dagger(x)~U^{-1}(x)
\label{pp}
\eea
which is the gauge transformation of the non-abelian phase in QCD under the non-abelian gauge transformation
as given by eq. (\ref{gtg}).

From eqs.  (\ref{ah}), (\ref{lkjn}) and (\ref{pp}) we find
\bea
{\cal P}e^{-ig \int_0^{\infty} d\lambda l\cdot { A}'^a(x+l\lambda)T^a }=U(x){\cal P}e^{-ig \int_0^{\infty} d\lambda l\cdot { A}^a(x+l\lambda)T^a },~~~~~~~U(x)=e^{igT^a\omega^a(x)}
\label{kkkj}
\eea
which is the gauge transformation of the non-abelian gauge link in QCD under the non-abelian gauge transformation
as given by eq. (\ref{gtg}).

From eqs. (\ref{lkjn}) and (\ref{kkkj}) we find that, under the non-abelian gauge transformation as given by eq.
(\ref{gtg}), the light-like Wilson line in QCD transforms as
\bea
{\cal P}e^{ig \int_{x_i}^{x_f} dx^\mu A'^a_\mu(x)T^a }=U(x_f)\left[{\cal P}e^{ig \int_{x_i}^{x_f} dx^\mu A^a_\mu(x)T^a }\right]U^{-1}(x_i),~~~~~~~U(x)=e^{igT^a\omega^a(x)}.
\label{ssma}
\eea

\section{ Emergence of Light-Like Wilson Line in QCD in the Proof of Factorization Theorem at High Energy Colliders }

Note that in \cite{nayak1f} we have shown that the light-like Wilson line in QCD naturally emerges when path integral formulation
is used to prove NRQCD factorization at all order in coupling constant in heavy quarkonium production.
Similarly, in \cite{nayak2f} we have shown that the light-like Wilson line in QCD naturally emerges when path integral formulation
is used to prove factorization of soft and collinear divergences of the gluon distribution function at high energy colliders
at all order in coupling constant. In this paper we will prove that the light-like Wilson line in QCD naturally emerges when
path integral formulation is used to prove factorization of soft and collinear divergences of the quark distribution function at
high energy colliders at all order in coupling constant.

The generating functional in the path integral method of QCD is given by \cite{muta,abbott}
\bea
Z[J,\eta,{\bar \eta}]=\int [dQ] [d{\bar \psi}] [d \psi ] ~{\rm det}(\frac{\delta \partial_\mu Q^{\mu a}}{\delta \omega^b})
~e^{i\int d^4x [-\frac{1}{4}{F^a}_{\mu \nu}^2[Q] -\frac{1}{2 \alpha} (\partial_\mu Q^{\mu a})^2+{\bar \psi} [i\gamma^\mu \partial_\mu -m +gT^a\gamma^\mu Q^a_\mu] \psi + J \cdot Q +{\bar \eta} \psi +  {\bar \psi} \eta]} \nonumber \\
\label{zfq}
\eea
where $J^{\mu a}(x)$ is the external source for the quantum gluon field $Q^{\mu a}(x)$ and ${\bar \eta}_i(x)$ is the external source for the
Dirac field $\psi_i(x)$ of the quark and
\bea
F^a_{\mu \nu}[Q]=\partial_\mu Q^a_\nu(x)-\partial_\nu Q^a_\mu(x)+gf^{abc}Q^b_\mu(x)Q^c_\nu(x),~~~~~~~~~{F^a}_{\mu \nu}^2[Q]={F}^{\mu \nu a}[Q]{F}^a_{\mu \nu}[Q].
\eea
The light-like quark traveling with light-like four-velocity $l^\mu$ produces SU(3) pure gauge potential $A^{\mu a}(x)$
at all the time-space position $x^\mu$ except at the position ${\vec x}$ perpendicular to the direction of motion
of the quark (${\vec l}\cdot {\vec x}=0$) at the time of closest approach \cite{collinssterman,nayakj,nayake}.
Hence the soft and collinear behavior of the non-perturbative
correlation function in QCD due to the presence of light-like Wilson line in QCD can be studied by using path integral formulation of
the background field method of QCD in the presence of SU(3) pure gauge background field \cite{nayak1f,nayak2f}.

Background field method of QCD was originally formulated by 't Hooft \cite{thooft} and later
extended by Klueberg-Stern and Zuber \cite{zuber,zuber1} and by Abbott \cite{abbott}.
This is an elegant formalism which can be useful to construct gauge invariant
non-perturbative green's functions in QCD. This formalism is also useful to study quark and gluon production from classical chromo field \cite{peter}
via Schwinger mechanism \cite{schw}, to compute $\beta$ function in QCD \cite{peskin}, to perform
calculations in lattice gauge theories \cite{lattice} and to study evolution of QCD
coupling constant in the presence of chromofield \cite{nayak}.

It can be mentioned here that in soft collinear effective theory
(SCET) \cite{scet} it is also necessary to use the idea of background fields \cite{abbott} to give well defined meaning to several
distinct gluon fields \cite{scet1}.

Note that a massive color source traveling at speed much less than speed of light
can not produce SU(3) pure gauge field \cite{collinssterman,nayakj,nayake}. Hence when one replaces light-like
Wilson line with massive Wilson line one expects the factorization of soft/infrared divergences to
break down. This is in conformation with the finding in \cite{nayaksterman1} which used the diagrammatic
method of QCD. In case of massive Wilson line in QCD the color transfer occurs and the factorization breaks
down. Note that in case of massive Wilson line there is no collinear divergences.

The generating functional in the path integral formulation of the background field method of QCD is
given by \cite{thooft,abbott,zuber}
\bea
&& Z[A,J,\eta,{\bar \eta}]=\int [dQ] [d{\bar \psi}] [d \psi ] ~{\rm det}(\frac{\delta G^a(Q)}{\delta \omega^b}) \nonumber \\
&& e^{i\int d^4x [-\frac{1}{4}{F^a}_{\mu \nu}^2[A+Q] -\frac{1}{2 \alpha}
(G^a(Q))^2+{\bar \psi} [i\gamma^\mu \partial_\mu -m +gT^a\gamma^\mu (A+Q)^a_\mu] \psi + J \cdot Q +{\bar \eta} \psi + {\bar \psi}\eta ]}
\label{azaqcd}
\eea
where the gauge fixing term is given by
\bea
G^a(Q) =\partial_\mu Q^{\mu a} + gf^{abc} A_\mu^b Q^{\mu c}=D_\mu[A]Q^{\mu a}
\label{ga}
\eea
which depends on the background field $A^{\mu a}(x)$ and
\bea
F_{\mu \nu}^a[A+Q]=\partial_\mu [A_\nu^a+Q_\nu^a]-\partial_\nu [A_\mu^a+Q_\mu^a]+gf^{abc} [A_\mu^b+Q_\mu^b][A_\nu^c+Q_\nu^c].
\eea
We have followed the notations of \cite{thooft,zuber,abbott} and accordingly we have
denoted the quantum gluon field by $Q^{\mu a}$ and the background field by $A^{\mu a}$.

Note that the gauge fixing term $\frac{1}{2 \alpha} (G^a(Q))^2$ in eq. (\ref{azaqcd}) [where $G^a(Q)$ is given by eq. (\ref{ga})]
is invariant for gauge transformation of $A_\mu^a$:
\bea
\delta A_\mu^a = gf^{abc}A_\mu^b\omega^c + \partial_\mu \omega^a,  ~~~~~~~({\rm type~ I ~transformation})
\label{typeI}
\eea
provided one also performs a homogeneous transformation of $Q_\mu^a$ \cite{zuber,abbott}:
\bea
\delta Q_\mu^a =gf^{abc}Q_\mu^b\omega^c.
\label{omega}
\eea
The gauge transformation of background field $A_\mu^a$ as given by eq. (\ref{typeI})
along with the homogeneous transformation of $Q_\mu^a$ in eq. (\ref{omega}) gives
\bea
\delta (A_\mu^a+Q_\mu^a) = gf^{abc}(A_\mu^b+Q_\mu^b)\omega^c + \partial_\mu \omega^a
\label{omegavbxn}
\eea
which leaves $-\frac{1}{4}{F^a}_{\mu \nu}^2[A+Q]$ invariant in eq. (\ref{azaqcd}).

For fixed $A_\mu^a$, {\it i.e.}, for
\bea
&&\delta A_\mu^a =0,  ~~~~~~~({\rm type~ II ~transformation})
\label{typeII}
\eea
the gauge transformation of $Q_\mu^a$ \cite{zuber,abbott}:
\bea
&&\delta Q_\mu^a = gf^{abc}(A_\mu^b + Q_\mu^b)\omega^c + \partial_\mu \omega^a
\label{omegaII}
\eea
gives eq. (\ref{omegavbxn}) which leaves $-\frac{1}{4}{F^a}_{\mu \nu}^2[A+Q]$ invariant in eq. (\ref{azaqcd}).

It is useful to remember that, unlike QED \cite{tucci}, finding an exact relation between the generating
functional $Z[J, \eta, {\bar \eta}]$ in QCD in eq. (\ref{zfq}) and the generating functional $Z[A, J, \eta, {\bar \eta}]$
in the background field method of QCD in eq. (\ref{azaqcd}) in the presence of SU(3) pure gauge background field is not easy.
The main difficulty is due to the gauge fixing terms which are different in both the cases. While the Lorentz (covariant) gauge fixing
term  $-\frac{1}{2 \alpha}(\partial_\mu Q^{\mu a})^2$ in eq. (\ref{zfq}) in QCD is independent of the background field
$A^{\mu a}(x)$, the background field gauge fixing term $-\frac{1}{2 \alpha}(G^a(Q))^2$ in eq. (\ref{azaqcd}) in the background field method
of QCD depends on the background field $A^{\mu a}(x)$ where $G^a(Q)$ is given by eq. (\ref{ga}) \cite{thooft,zuber,abbott}.
Hence in order to study non-perturbative correlation function
in the background field method of QCD in the presence of SU(3) pure gauge background
field we proceed as follows.

By changing $Q \rightarrow Q-A$ in eq. (\ref{azaqcd}) we find
\bea
&& Z[A,J,\eta,{\bar \eta}]=e^{-i \int d^4x J \cdot A }\int [dQ] [d{\bar \psi}] [d \psi ]
 \nonumber \\
&&~{\rm det}(\frac{\delta G^a_f(Q)}{\delta \omega^b}) ~ e^{i\int d^4x [-\frac{1}{4}{F^a}_{\mu \nu}^2[Q] -\frac{1}{2 \alpha}
(G^a_f(Q))^2+ J \cdot Q +{\bar \psi} [i\gamma^\mu \partial_\mu -m +gT^a\gamma^\mu Q^a_\mu] \psi
+{\bar \eta} \psi + {\bar \psi}\eta] }
\label{zaqcd1}
\eea
where the gauge fixing term from eq. (\ref{ga}) becomes
\bea
G_f^a(Q) =\partial_\mu Q^{\mu a} + gf^{abc} A_\mu^b Q^{\mu c} - \partial_\mu A^{\mu a}=D_\mu[A] Q^{\mu a} - \partial_\mu A^{\mu a},
\label{gfa}
\eea
and eq. (\ref{omega}) [by using eq. (\ref{typeI}), type I transformation \cite{zuber,abbott}] becomes
\bea
&&\delta Q^a_\mu = gf^{abc}Q_\mu^b\omega^c+ \partial_\mu \omega^a.
\label{theta}
\eea
The eqs. (\ref{gfa}) and (\ref{theta}) can also be derived by using type II transformation which can be seen as follows.
By changing $Q \rightarrow Q-A$ in eq. (\ref{azaqcd}) we find eq. (\ref{zaqcd1})
where the gauge fixing term from eq. (\ref{ga}) becomes eq. (\ref{gfa})
and eq. (\ref{omegaII}) [by using eq. (\ref{typeII})] becomes eq. (\ref{theta}).
Hence we obtain eqs. (\ref{zaqcd1}), (\ref{gfa}) and (\ref{theta}) whether we use
the type I transformation or type II transformation. Hence we find that we will obtain the same
eq. (\ref{zaqcd1cz3}) whether we use the type I transformation or type II transformation.

The equation
\bea
Q'^a_\mu(x)= Q^a_\mu(x) +gf^{abc}\omega^c(x) Q_\mu^b(x) + \partial_\mu \omega^a(x)
\label{thetaav}
\eea
in eq. (\ref{theta}) is valid for infinitesimal transformation ($\omega << 1$) which is obtained from the
finite equation
\bea
T^aQ'^a_\mu(x) = U(x)T^aQ^a_\mu(x) U^{-1}(x)+\frac{1}{ig}[\partial_\mu U(x)] U^{-1}(x),~~~~~~~~~~~U(x)=e^{igT^a\omega^a(x)}.
\label{ftgrm}
\eea

Simplifying infinite numbers of non-commuting terms in eq. (\ref{ftgrm}) [by using eq. (\ref{utuinv}) and \cite{nayakj}] we find that
\bea
{Q'}_\mu^a(x) = [e^{gM(x)}]_{ab}Q_\mu^b(x) ~+ ~[\frac{e^{gM(x)}-1}{gM(x)}]_{ab}~[\partial_\mu \omega^b(x)],~~~~~~~~~~~M_{ab}(x)=f^{abc}\omega^c(x).
\label{teq}
\eea

Changing the variables of integration from unprimed to primed variables in eq. (\ref{zaqcd1}) we find
\bea
&& Z[A,J,\eta,{\bar \eta}]
=e^{-i \int d^4x J \cdot A }\int [dQ'][d{\bar \psi}'] [d \psi' ] \nonumber \\
&&~{\rm det}(\frac{\delta G^a_f(Q')}{\delta \omega^b}) ~ e^{i\int d^4x [-\frac{1}{4}{F^a}_{\mu \nu}^2[Q'] -\frac{1}{2 \alpha}
(G^a_f(Q'))^2+ J \cdot Q' +{\bar \psi}' [i\gamma^\mu \partial_\mu -m +gT^a\gamma^\mu Q'^a_\mu] \psi'
+{\bar \eta} \psi' + {\bar \psi}'\eta] }.
\label{zaqcd1b}
\eea
This is because a change of variables from unprimed to primed variables does not change the value of the
integration.

Under the finite transformation, using eq. (\ref{teq}), we find
\bea
&& [dQ'] =[dQ] ~{\rm det} [\frac{\partial {Q'}^a}{\partial Q^b}] = [dQ] ~{\rm det} [[e^{gM(x)}]]=[dQ] {\rm exp}[{\rm Tr}({\rm ln}[e^{gM(x)}])]=[dQ]
\label{dqa}
\eea
where we have used (for any matrix $H$)
\bea
{\rm det}H={\rm exp}[{\rm Tr}({\rm ln}H)].
\eea

The fermion field transforms as
\bea
\psi'(x)=e^{igT^a\omega^a(x)}\psi(x).
\label{phg3}
\eea
Using eqs. (\ref{teq}) and (\ref{phg3}) we find
\bea
&&[d{\bar \psi}'] [d \psi' ]=[d{\bar \psi}] [d \psi ],~~~~~~{\bar \psi}' [i\gamma^\mu \partial_\mu -m +gT^a\gamma^\mu Q'^a_\mu] \psi'={\bar \psi} [i\gamma^\mu \partial_\mu -m +gT^a\gamma^\mu Q^a_\mu]\psi,\nonumber \\
&&{F^a}_{\mu \nu}^2[Q']={F^a}_{\mu \nu}^2[Q].
\label{psa}
\eea

Using eqs. (\ref{dqa}) and (\ref{psa}) in eq. (\ref{zaqcd1b}) we find
\bea
&& Z[A,J,\eta,{\bar \eta}]
=e^{-i \int d^4x J \cdot A }\int [dQ] [d{\bar \psi}] [d \psi ] \nonumber \\
&&~{\rm det}(\frac{\delta G^a_f(Q')}{\delta \omega^b}) ~ e^{i\int d^4x [-\frac{1}{4}{F^a}_{\mu \nu}^2[Q] -\frac{1}{2 \alpha}
(G^a_f(Q'))^2+ J \cdot Q' +{\bar \psi} [i\gamma^\mu \partial_\mu -m +gT^a\gamma^\mu Q^a_\mu] \psi
+{\bar \eta} \psi' + {\bar \psi}'\eta] }.
\label{zaqcd1cv}
\eea
From eq. (\ref{gfa}) we find
\bea
G_f^a(Q') =\partial_\mu Q^{' \mu a} + gf^{abc} A_\mu^b Q^{' \mu c} - \partial_\mu A^{\mu a}.
\label{gfap}
\eea
By using eqs. (\ref{teq}) and (\ref{pg4}) in eq. (\ref{gfap}) we find
\bea
&&G_f^a(Q') =\partial^\mu [[e^{gM(x)}]_{ab}Q_\mu^b(x) ~+ ~[\frac{e^{gM(x)}-1}{gM(x)}]_{ab}~[\partial_\mu \omega^b(x)]]\nonumber \\
&&+ gf^{abc} [\partial^\mu \omega^e(x)\left[\frac{e^{gM(x)}-1}{gM(x)}\right]_{be}] [[e^{gM(x)}]_{cd}Q_\mu^d(x) ~+ ~[\frac{e^{gM(x)}-1}{gM(x)}]_{cd}~[\partial_\mu \omega^d(x)]]\nonumber \\
&&- \partial_\mu [\partial^\mu \omega^b(x)\left[\frac{e^{gM(x)}-1}{gM(x)}\right]_{ab}]
\label{gfapa}
\eea
which gives
\bea
&&G_f^a(Q') =\partial^\mu [[e^{gM(x)}]_{ab}Q_\mu^b(x)]\nonumber \\
&&+ gf^{abc} [\partial^\mu \omega^e(x)\left[\frac{e^{gM(x)}-1}{gM(x)}\right]_{be}] [[e^{gM(x)}]_{cd}Q_\mu^d(x) ~+ ~[\frac{e^{gM(x)}-1}{gM(x)}]_{cd}~[\partial_\mu \omega^d(x)]].
\label{gfapb}
\eea
From eq. (\ref{gfapb}) we find
\bea
&&G_f^a(Q') =\partial^\mu [[e^{gM(x)}]_{ab}Q_\mu^b(x)]+ gf^{abc} [\partial^\mu \omega^e(x)\left[\frac{e^{gM(x)}-1}{gM(x)}\right]_{be}] [[e^{gM(x)}]_{cd}Q_\mu^d(x)]
\label{gfapc}
\eea
which gives
\bea
&&G_f^a(Q') = [e^{gM(x)}]_{ab}\partial^\mu Q_\mu^b(x)\nonumber \\
&&+Q_\mu^b(x)\partial^\mu [[e^{gM(x)}]_{ab}]+  [\partial^\mu \omega^e(x)\left[\frac{e^{gM(x)}-1}{gM(x)}\right]_{be}]gf^{abc} [[e^{gM(x)}]_{cd}Q_\mu^d(x)].
\label{gfapd}
\eea
From \cite{nayakj} we find
\bea
\partial^\mu [e^{igT^a\omega^a(x)}]_{ij}=ig[\partial^\mu \omega^b(x)]\left[\frac{e^{gM(x)}-1}{gM(x)}\right]_{ab}T^a_{ik}[e^{igT^c\omega^c(x)}]_{kj},~~~~~~~~~M_{ab}(x)=f^{abc}\omega^c(x)\nonumber \\
\label{pg4j}
\eea
which in the adjoint representation of SU(3) gives (by using $T^a_{bc}=-if^{abc}$)
\bea
[\partial^\mu e^{gM(x)}]_{ad}=[\partial^\mu \omega^e(x)]\left[\frac{e^{gM(x)}-1}{gM(x)}\right]_{be}gf^{bac}[e^{M(x)}]_{cd},~~~~~~~~~M_{ab}(x)=f^{abc}\omega^c(x).
\label{pg4k}
\eea
Using eq. (\ref{pg4k}) in (\ref{gfapd}) we find
\bea
&&G_f^a(Q') = [e^{gM(x)}]_{ab}\partial^\mu Q_\mu^b(x)
\label{gfape}
\eea
which gives
\bea
(G_f^a(Q'))^2 = (\partial_\mu Q^{\mu a}(x))^2.
\label{gfapf}
\eea
Since for $n \times n$ matrices $A$ and $B$ we have
\bea
{\rm det}(AB)=({\rm det}A)({\rm det} B)
\eea
we find by using eq. (\ref{gfape}) that
\bea
&&{\rm det} [\frac{\delta G_f^a(Q')}{\delta \omega^b}] ={\rm det}
[\frac{ \delta [[e^{gM(x)}]_{ac}\partial^\mu Q_\mu^c(x)]}{\delta \omega^b}]={\rm det}[
[e^{gM(x)}]_{ac}\frac{ \delta (\partial^\mu Q_\mu^c(x))}{\delta \omega^b}]\nonumber \\
&&=\left[{\rm det}[
[e^{gM(x)}]_{ac}]\right]~\left[{\rm det}[\frac{ \delta (\partial^\mu Q_\mu^c(x))}{\delta \omega^b}]\right]={\rm exp}[{\rm Tr}({\rm ln}[e^{gM(x)}])]~{\rm det}[\frac{ \delta (\partial_\mu Q^{\mu a}(x))}{\delta \omega^b}]\nonumber \\
&&={\rm det}[\frac{ \delta (\partial_\mu Q^{\mu a}(x))}{\delta \omega^b}].
\label{gqp4a}
\eea
Using eqs. (\ref{gfapf}) and (\ref{gqp4a}) in eq. (\ref{zaqcd1cv}) we find
\bea
&& Z[A,J,\eta,{\bar \eta}]=e^{-i \int d^4x J \cdot A }\int [dQ] [d{\bar \psi}] [d \psi ] \nonumber \\
&&~{\rm det}[\frac{ \delta (\partial_\mu Q^{\mu a}(x))}{\delta \omega^b}] ~ e^{i\int d^4x [-\frac{1}{4}{F^a}_{\mu \nu}^2[Q] -\frac{1}{2 \alpha}
(\partial_\mu Q^{\mu a})^2+ J \cdot Q'+{\bar \psi} [i\gamma^\mu \partial_\mu -m +gT^a\gamma^\mu Q^a_\mu] \psi
+{\bar \eta} \psi' + {\bar \psi}'\eta] }.
\label{zaqcd1cz3}
\eea
From eqs. (\ref{pg4}) and (\ref{teq}) we find
\bea
{Q'}_\mu^a(x) -A_\mu^a(x)= [e^{gM(x)}]_{ab}Q_\mu^b(x),~~~~~~~~~~~M_{ab}(x)=f^{abc}\omega^c(x).
\label{tej}
\eea
Note that eqs. (\ref{zaqcd1cz3}) and (\ref{tej}) are valid whether we use type I
transformation [eqs. (\ref{typeI}) and (\ref{omega})] or type II transformation [eqs. (\ref{typeII}) and (\ref{omegaII})].

Since we have used eq. (\ref{gtg}) to study the gauge transformation of the Wilson line in QCD we will use
type I transformation, see eqs. (\ref{typeI}) and (\ref{omega}), in the rest of the paper which
gives for finite transformation \cite{abbott,zuber}
\bea
J'^a_\mu(x)=[e^{gM(x)}]_{ab}J_\mu^b(x),~~~~~~~~~~~M_{ab}(x)=f^{abc}\omega^c(x).
\label{jpr}
\eea
From eqs. (\ref{zaqcd1cz3}), (\ref{tej}) and (\ref{jpr}) we find
\bea
&& Z[A,J',\eta,{\bar \eta}]
~=~\int [dQ] [d{\bar \psi}] [d \psi ] \nonumber \\
&&~{\rm det}[\frac{ \delta (\partial_\mu Q^{\mu a}(x))}{\delta \omega^b}] ~ e^{i\int d^4x [-\frac{1}{4}{F^a}_{\mu \nu}^2[Q] -\frac{1}{2 \alpha}
(\partial_\mu Q^{\mu a})^2+ J \cdot Q +{\bar \psi} [i\gamma^\mu \partial_\mu -m +gT^a\gamma^\mu Q^a_\mu]\psi
+{\bar \eta} \psi' + {\bar \psi}'\eta] }.
\label{zaqcd1cz}
\eea
Under the non-abelian gauge transformation the fermion source transforms as \cite{zuber,abbott}
\bea
\eta'(x)=e^{igT^a\omega^a(x)}\eta(x).
\label{fqcd1}
\eea

From eqs. (\ref{phg3}) and (\ref{fqcd1}) we find
\bea
{\bar \eta}' \psi'={\bar \eta} \psi,~~~~~~~~{\bar \psi}'\eta'={\bar \psi}\eta
\label{ppl}
\eea
which gives from eq. (\ref{zaqcd1cz})
\bea
&& Z[A,J',\eta',{\bar \eta}']
~=~\int [dQ] [d{\bar \psi}] [d \psi ] \nonumber \\
&&~{\rm det}[\frac{ \delta (\partial_\mu Q^{\mu a}(x))}{\delta \omega^b}] ~ e^{i\int d^4x [-\frac{1}{4}{F^a}_{\mu \nu}^2[Q] -\frac{1}{2 \alpha}
(\partial_\mu Q^{\mu a})^2+ J \cdot Q +{\bar \psi} [i\gamma^\mu \partial_\mu -m +gT^a\gamma^\mu Q^a_\mu] \psi
+{\bar \eta} \psi + {\bar \psi}\eta] }.
\label{zaqcd1c}
\eea
Hence from eqs. (\ref{zaqcd1c}) and (\ref{zfq}) we find that in QCD
\bea
Z[J,\eta,{\bar \eta}]=Z[A,J',\eta',{\bar \eta}']
\label{fqcd}
\eea
when the  background field $A^{\mu a}(x)$ is the SU(3) pure gauge as given by eq. (\ref{gtqcd}).
The corresponding relation in QED is given by
\bea
Z[J,\eta,{\bar \eta}]=Z[A,J,\eta',{\bar \eta}']
\label{fqed}
\eea
when the background field $A^\mu(x)$ is the U(1) pure gauge as given by eq. (\ref{pgqed}).
Note that, unlike eq. (\ref{fqcd}) in QCD there is no $J'$ in eq. (\ref{fqed}) in QED because while the
(quantum) gluon directly interacts with classical chromo-electromagnetic field the (quantum) photon does
not directly interact with classical electromagnetic field.

The non-perturbative correlation function of the type $<0|{\bar \psi}(x) \psi(x')|0>$ in QCD is
given by \cite{tucci}
\bea
<0|{\bar \psi}(x) \psi(x')|0>=\frac{\delta}{\delta \eta(x)}\frac{\delta}{\delta {\bar \eta}(x')}Z[J,\eta,{\bar \eta}]|_{J=\eta={\bar \eta}=0}.
\label{fqcd2}
\eea
Similarly the non-perturbative correlation function of the type $<0|{\bar \psi}(x) \psi(x')|0>_A$
in the background field method of QCD is given by \cite{tucci}
\bea
<0|{\bar \psi}(x) \psi(x')|0>_A=\frac{\delta}{\delta \eta(x)}\frac{\delta}{\delta {\bar \eta}(x')}Z[A,J,\eta,{\bar \eta}|_{J=\eta={\bar \eta}=0}.
\label{fqcd3}
\eea
When background field $A^{\mu a}(x)$ is the SU(3) pure gauge as given by eq. (\ref{gtqcd}) we find from eqs. (\ref{fqcd}), (\ref{fqcd2}),
(\ref{fqcd3}), (\ref{jpr}) and (\ref{fqcd1}) that
\bea
<0|{\bar \psi}(x) \psi(x')|0>=<0|{\bar \psi}(x) \Phi(x) \Phi^\dagger(x') \psi(x')|0>_A
\label{fqcd4}
\eea
which proves factorization of soft and collinear divergences at all order in coupling constant in QCD
where [see eq. (\ref{lkjn}) and \cite{nayak1f,nayak2f}]
\bea
\Phi(x)=~{\cal P}{\rm exp}[-igT^a \int_0^{\infty} d\lambda l\cdot { A}^a(x+l\lambda) ]=e^{igT^a\omega^a(x)}
\label{fqcdf}
\eea
is the non-abelian phase or the gauge-link in QCD.

From eq. (\ref{fqcd4}) we find that the correct definition of the quark distribution function at high energy colliders which is
consistent with the number operator interpretation of the quark and is gauge invariant and is consistent with the
factorization theorem in QCD is given by
\bea
&& f_{q/P}(x)= \frac{1}{4\pi}\int dy^- e^{-ix{P}^+ y^- }\times <P| {\bar \psi}(0,y^-,0_T) \gamma^+
[{\cal P}{\rm exp}[igT^a\int_0^{y^-}dz^-A^{+a}(0,z^-,0_T)]]\psi(0)|P>\nonumber \\
\label{gpdffpi}
\eea
which is valid in covariant gauge, in light-cone gauge, in general axial gauges, in general non-covariant gauges and in
general Coulomb gauge etc. respectively \cite{nayak2f}. In eq. (\ref{gpdffpi}) the $\psi(x)$ is the Dirac field of the quark and
$A^{\mu a}(x)$ is the SU(3) pure gauge background field as given by eq. (\ref{gtqcd}).

Hence we find from eq. (\ref{gpdffpi}) and from \cite{nayak1f,nayak2f}
that the light-like Wilson line in QCD naturally emerges when path integral formulation of QCD is used to
prove factorization of soft and collinear divergences at all order in coupling constant in QCD
processes at high energy colliders.

\section{ Conclusions }
Unlike the Wilson line in QED the Wilson line in QCD contains path ordering.
In this paper we get rid of the
path ordering in the light-like Wilson line in QCD by simplifying all the infinite number
of non-commuting terms in the SU(3) pure gauge. We have proved that the
light-like Wilson line in QCD naturally emerges when path integral formulation of QCD is used to prove factorization of
soft and collinear divergences at all order in coupling constant in QCD processes at high energy colliders.

\end{document}